\documentclass[12pt]{article}
\usepackage{graphicx}
\usepackage{amssymb}

\newcommand{\be}{\begin{equation}}
\newcommand{\ee}{\end{equation}}
\newcommand{\bea}{\begin{eqnarray}}
\newcommand{\eea}{\end{eqnarray}}
\newcommand{\beq}{\begin{equation}}
\newcommand{\eeq}{\end{equation}}
\newcommand{\ba}{\begin{array}}
\newcommand{\ea}{\end{array}}
\newcommand{\beqa}{\begin{eqnarray}}
\newcommand{\eeqa}{\end{eqnarray}}

\newcommand{\cO}{{\cal O}}

\newcommand{\lsim}{\stackrel{<}{_\sim}}

\def \text{\mathrm}

\newcommand{\LL}{{\mbox{\scriptsize LL}}}
\newcommand{\LR}{{\mbox{\scriptsize LR}}}
\newcommand{\RL}{{\mbox{\scriptsize RL}}}
\newcommand{\RR}{{\mbox{\scriptsize RR}}}

\begin{document}

\begin{center}

\vspace{1cm}

{\LARGE{\bf The B -- TAU FCNC connection in SUSY Unified Theories\
\footnote{Based on talks given at: 
{\em DIF06, International Workshop on discoveries in flavour physics at e+e- colloders},
Laboratori Nazionali di Frascati (Italy), February 28- March 03, 2006;
{\em XLIst Rencontres de Moriond}, La Thuile, 5-11 March 2006;
{\em CORFU2005, Corfu Summer Institute on EPP}, Corfu, Greece, September 4-26, 2005.
}}}

\vspace{1.0cm}

{\large{\sc A.~Masiero$^1$ and P.~Paradisi$^2$}}

\vspace{0.6cm}

$^1$Univ. of Padova and INFN, Padova, Italy \\
$^2$Department of Physics, Technion-Israel Institute of Technology, \\
{\em Technion City, 32000 Haifa, Israel}\\

\end{center}

\vspace{1cm}

{\abstract
\noindent
In the context of SUSY grand unification a link is established between the hadronic 
and leptonic soft breaking sectors. Such relation is here exploited in particular 
for FCNC processes in B physics. It is shown how bounds on leptonic FCNC involving 
the third generation translate into constraints on FC B decays. 
In the second part of the contribution we show that tests of lepton universality 
in K and B decays can represent an interesting handle to obtain relevant information 
on the amount of FCNC in the second and third fermion generation.}

\newpage

\pagestyle{plain}

\section{Grand Unification of Quark and Lepton FCNCs}
Supersymmetry (SUSY) breaking (SB) remains one of the biggest issues
in physics beyond the Standard Model (SM). In spite of various
proposals, we still miss a realistic and theoretically satisfactory
model of SB.

Flavor violating processes have been instrumental in guiding us
towards consistent SB models.  Indeed, even in the absence of a
well-defined SB mechanism and, hence, without a precise knowledge of
the SUSY lagrangian at the electroweak scale, it is still possible to
make use of the FCNC bounds to infer relevant constraints on the part
of the SUSY soft breaking sector related to the sfermion mass matrices.

The model-independent method which is adopted is
the so-called Mass-Insertion approximation (MIA). In
this approach, the experimental limits lead to upper bounds on the
parameters (or combinations of) $\delta_{ij}^f \equiv
\Delta^f_{ij}/m_{\tilde{f}}^2$; where $\Delta^f_{ij}$ is the
flavour-violating off-diagonal entry appearing in the $f = (u,d,l)$
sfermion mass matrices and $m_{\tilde{f}}^2$ is the average sfermion
mass. The mass-insertions include the LL/LR/RL/RR types, according to
the chirality of the corresponding SM fermions.

Detailed bounds on
the individual $\delta$s have been derived by considering limits from
various FCNC processes \cite{gabbiani}.  As long as one remains within
the simple picture of the Minimal Supersymmetric Standard Model
(MSSM), where quarks and leptons are unrelated, the hadronic and
leptonic FCNC processes yield separate bounds on the corresponding
$\delta^q$'s and $\delta^l$'s, respectively.

The situation changes when one embeds the MSSM within a Grand Unified
Theory (GUT).  In a SUSY GUT, quarks and leptons sit in same
multiplets and are transformed into each other through GU symmetry
transformations. If the supergravity lagrangian, and, in particular,
its K\"ahler function are present at a scale larger than the GUT
breaking scale, they have to fully respect the underlying gauge
symmetry which is the GU symmetry itself.  The subsequent SUSY
breaking will give rise to the usual soft breaking terms in the
lagrangian. In particular, if the mediation mechanism responsible for
the transmission of the SUSY breaking to the visible sector is
gravitational, the sfermion mass matrices, whose structure is dictated
by the K\"ahler potential, will have to respect the
underlying GU symmetry. Hence we expect quark-lepton correlations
among entries of the sfermion mass matrices \cite{ourprl}.
In other words, the quark-lepton unification seeps also into the
SUSY breaking soft sector.

Imposition of a GU symmetry on the $\mathcal{L}_{\rm soft}$ entails
relevant implications at the weak scale. This is because the flavour
violating (FV) mass-insertions do not get strongly renormalized
through RG scaling from the GUT scale to the weak scale in the absence
of new sources of flavor violation. On the other hand, if such new
sources are present, for instance due to the presence of new neutrino
Yukawa couplings in SUSY GUTs with a seesaw mechanism for neutrino
masses, then one can compute the RG-induced effects in terms of these
new parameters. Hence, the correlations between hadronic and leptonic
flavor violating MIs survive at the weak scale to a good
approximation. As for the flavor conserving (FC) mass insertions 
(i.e., the diagonal entries of the sfermion mass matrices), they get
strongly renormalized but in a way which is RG computable.

The connection between quark and lepton $\delta$ parameters can have
significant implications on flavor phenomenology \cite{ourprl,othercorrelations}. 
Indeed, using these relations, a quark $\delta$ parameter can be probed in a leptonic
process or vice versa. In this way, it is possible that constraints in
one sector are converted to the other sector where previously only
weaker or perhaps even no bounds existed. A thorough analysis along 
these lines has been performed by our group and is going to appear very soon 
\cite{ourfuture}. This extends and quantitatively accomplishes the research 
project outlined in our previous work \cite{ourprl}.
Here we present a limited selection of such results, in particular concerning 
B physics.

To be specific, we concentrate on the SUSY $SU(5)$ framework and
derive all the relations between squark and sleptonic mass
insertions. We then study the impact of the limit from $\tau \rightarrow
\mu\, \gamma$ on the $b \to s $ transition observables, such as
$A_{CP} (B \to \phi K_s)$.

The soft terms are assumed to be generated at some scale above
$M_{GUT}$.  Note that even assuming complete universality of the soft
breaking terms at $M_{Planck}$, as in mSUGRA, the RG effects to
$M_{GUT}$ will induce flavor off-diagonal entries at the GUT scale
\cite{fbam,barbieri}.  Hence we assume generic flavor violating entries to
be present in the sfermion matrices at the GUT scale.
Let us consider the scalar soft breaking sector of the MSSM:

\bea
\label{smsoft}
- {\cal L}_{\rm soft}\!\!\!\!\! &=& \!\!\!\!m_{Q_{ii}}^2 \tilde{Q}_i^\dagger \tilde{Q}_i 
+ m_{u^c_{ii}}^2 \tilde{u^c}_i^\star \tilde{u^c}_i + m^2_{e^c_{ii}} 
\tilde{e^c}_i^\star \tilde{e^c}_i 
 +   m^2_{d^c_{ii}} \tilde{d^c}^\star_i 
\tilde{d^c}_i + m_{L_{ii}}^2 \tilde{L}_i^\dagger \tilde{L}_i 
\nonumber \\ 
&+&\!\!\!\! 
m^2_{H_1} H^\dagger_1 H_1 
+  m^2_{H_2} H_2^\dagger H_2  \nonumber \\ 
&+& \!\!\!\!
A^u_{ij}~\tilde{Q}_i \tilde{u^c}_j H_2 + A^d_{ij}~
\tilde{Q}_i \tilde{d^c}_j H_1 + A^e_{ij}~
\tilde{L}_i \tilde{e^c}_j H_1 + 
(\Delta^{l}_{ij})_{\LL} \tilde{L}_{i}^\dagger \tilde{L}_{j}  + 
(\Delta^e_{ij})_{\RR} \tilde{e^c}_i^\star \tilde{e^c}_j  \nonumber \\ 
&+& \!\!\!\!
(\Delta^q_{ij})_{\LL} \tilde{Q}_i^\dagger \tilde{Q}_j  + 
(\Delta^u_{ij})_{\RR} \tilde{u^c}_i^\star \tilde{u^c}_j  + 
(\Delta^d_{ij})_{\RR} \tilde{d^c}_i^\star \tilde{d^c}_j  
+ (\Delta^e_{ij})_{\LR} \tilde{e_L}_i^\star \tilde{e^c}_j  
+ (\Delta^u_{ij})_{\LR} \tilde{u_L}_i^\star \tilde{u^c}_j \nonumber \\ 
&+& \!\!\!\!
(\Delta^d_{ij})_{\LR} \tilde{d_L}_i^\star \tilde{d^c}_j + \ldots 
\eea
where we have used the standard notation for the MSSM fields and have 
explicitly written down the various $\Delta$ parameters.

Consider that $SU(5)$ be the relevant symmetry at the scale where the
above soft terms firstly show up.  Then, taking into account that
matter is organized into the SU(5) representations ${\bf 10}~
=~(q,u^c,e^c)$ and ${\bf \overline 5}~ = ~(l,d^c)$, one obtains the
following relations
\bea
\label{matrel1}
m^2_{Q} = m^2_{\tilde{e^c}} = m^2_{\tilde{u^c}} = m^2_{\bf 10} \\
\label{matrel2}
m^2_{\tilde{d^c}} = m^2_{L} = m^2_{\bar{\bf \overline 5}} \\
\label{trirel}
A^e_{ij} = A^d_{ji}\, .
\eea
Eqs.~(\ref{matrel1}, \ref{matrel2}, \ref{trirel}) are matrices in flavor
space.  These equations lead to relations between the slepton and squark 
flavor violating off-diagonal entries $\Delta_{ij}$. These are: 
\bea
\label{cdeltas1}
(\Delta^u_{ij})_{\LL} = (\Delta^u_{ij})_{\RR} = (\Delta^d_{ij})_{\LL} =
(\Delta^l_{ij})_{\RR} \\
\label{cdeltas3}
(\Delta^d_{ij})_{\RR} = (\Delta^l_{ij})_{\LL} \\
\label{cdeltas4}
(\Delta^d_{ij})_{\LR} = (\Delta^l_{ji})_{\LR} = (\Delta^l_{ij})_{\RL}^\star.
\eea
These GUT correlations among hadronic and leptonic scalar soft terms
are summarized in table~\ref{tb0}. Assuming that
no new sources of flavor structure are present from the $SU(5)$ scale
down to the electroweak scale, apart from the usual SM CKM one, one
infers the relations in the first column of table~\ref{tb0} at low
scale.  
 \begingroup
 \begin{table}
 \begin{center}
 \begin{tabular*}{0.8\textwidth}{@{\extracolsep{\fill}}||c|c|c||}
 \hline\hline
&Weak-scale & 
GUT scale
\\[0.2pt] 
 \hline
 (1) & $(\delta^u_{ij})_{\RR}~ \approx~ (m_{e^c}^2/ m_{u^c}^2)~ 
 (\delta^l_{ij})_{\RR}$ & $m^2_{{u^c}_0} ~=~ m^2_{{e^c}_0}$ \\
 \hline
 (2) & 
 $(\delta^q_{ij})_{\LL}~ \approx~(m_{e^c}^2/ m_{Q}^2)~ (\delta^l_{ij})_{\RR}$ &
 $m^2_{{Q}_0} ~=~ m^2_{{e^c}_0}$ \\ 
 \hline 
 (3) &
 $(\delta^d_{ij})_{\RR} ~\approx~ (m_{L}^2/ m_{d^c}^2)~ (\delta^l_{ij})_{\LL}$ &
 $m^2_{{d^c}_0} ~=~ m^2_{{L}_0}$ \\ 
 \hline 
 (4) &
 $(\delta^d_{ij})_{\LR}\!\approx\!(m_{L}^2 /m_{Q}^2) 
 (m_b/ m_\tau) (\delta^l_{ij})_{\RL}^\star$  & 
 $A^e_{{ij}_0} = A^d_{{ji}_0}$
\\\hline\hline
 \end{tabular*}
 \end{center}
 \caption{Links between various transitions between up-type, down-type quarks 
 and charged leptons for SU(5). The suffix `0' implies GUT scale parameters.}
 \label{tb0}
 \end{table}
 \endgroup
Two comments are in order when looking at table~\ref{tb0}. 
First, the boundary conditions on the sfermion masses at the GUT scale 
(last column in table~\ref{tb0}) imply that the squark masses are
\textit{always} going to be larger at the weak scale compared to the
slepton masses. As a second remark, notice that some of the relations 
between hadronic and leptonic $\delta$ MIs in table~\ref{tb0} exhibit
opposite ``chiralities", i.e. LL insertions are related to RR ones and
vice-versa.  This stems from the arrangement of the different fermion
chiralities in $SU(5)$ five- and ten-plets (as it clearly appears
from the final column in table~\ref{tb0}). This restriction can easily
be overcome if we move from $SU(5)$ to left-right symmetric unified
models like SO(10) or the Pati-Salam (PS) case.

\begin{figure}
\includegraphics[width=17.0cm]{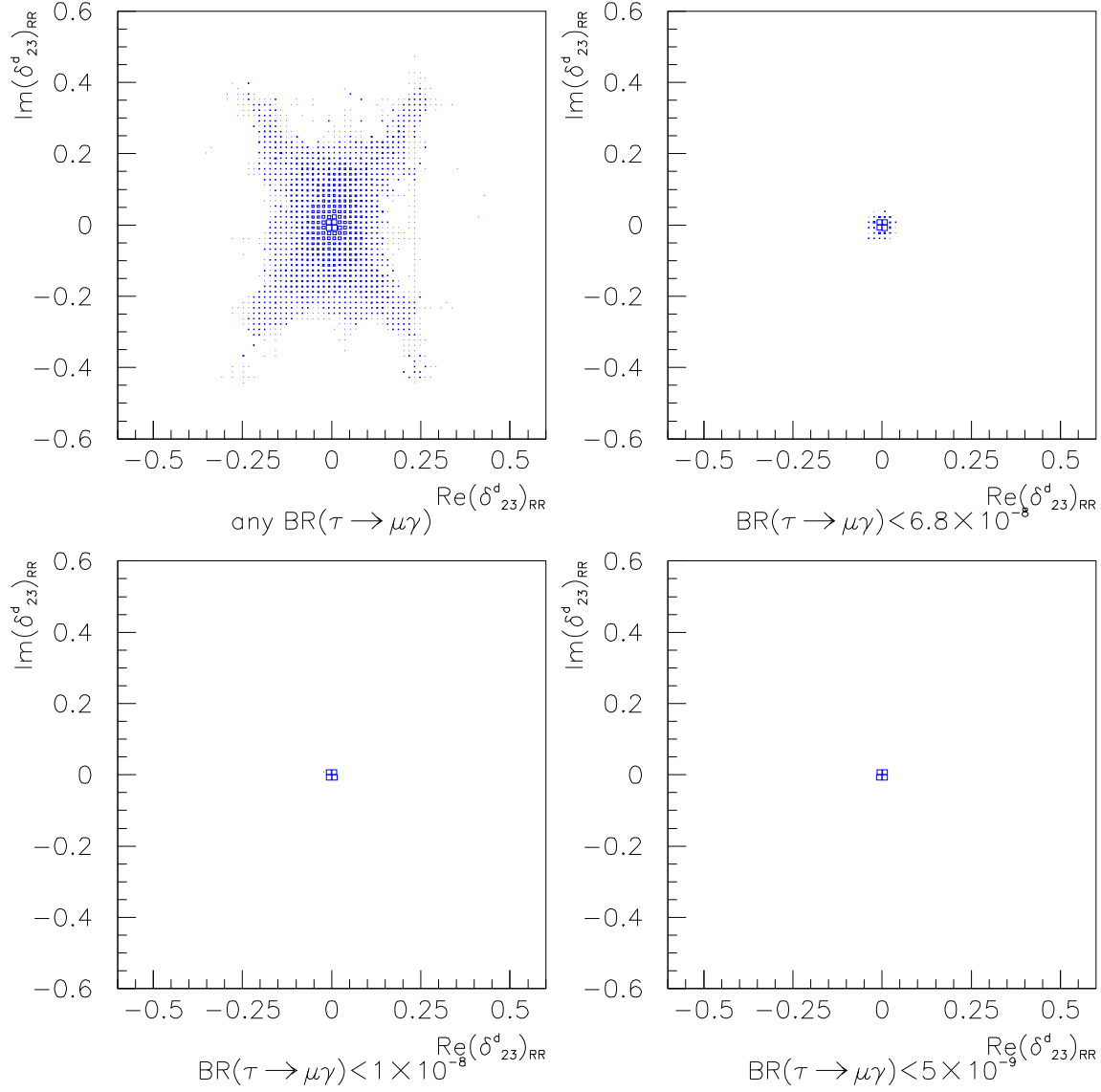}
\caption{
Allowed regions in the 
Re$(\delta^d_{23})_{RR}$--Im$(\delta^d_{23})_{RR}$ plane
for different values of $Br(\tau \to \mu \gamma)$.
Constraints from $B \to X_s \gamma$, $BR(B \to X_s \ell^+ \ell^-)$,
$B \to \phi K_s$ and $\Delta M_s$ have been used.
}
\label{fig:combi}
\end{figure}

In Fig.\ref{fig:combi}, we plot the probability density in
the Re$(\delta^d_{23})_{RR}$--Im$(\delta^d_{23})_{RR}$ plane for
different upper bounds on $BR(\tau \to \mu \, \gamma)$.
NLO branching ratios and CP asymmetries
for $B \to X_s \gamma$, $B \to \phi K_s$, $BR(B \to X_s \ell^+\ell^-)$ 
and $\Delta M_s$ were considered.
As shown in Fig.\ref{fig:combi}, the bound on $(\delta^d_{23})_{RR}$ 
induced by $BR(\tau \to \mu \, \gamma)$ is already at present much stronger 
than the bounds from hadronic processes, reducing considerably the room left 
for SUSY effects in $B$ decays.

Note that making use of the relation (3) with $|(\delta^l_{23})_{LL}|~ <~ 1
$, implies $\vert (\delta^d_{23})_{RR} \vert \lsim 0.5$ as the ratio
$(m_{L}^2/m_{d^c}^2)$ varies roughly between $(0.2 - 0.5)$ at the weak
scale, for the chosen high scale boundary conditions.  The effect on
$(\delta^d_{23})_{RR}$ of the upper bound on $BR(\tau \to \mu \,
\gamma)$ is dramatic already with the present experimental value.

\section{Lepton Universality in $K\rightarrow \ell\nu$}

High precision electroweak tests represent a powerful tool to probe 
the SM and, hence, to constrain or obtain indirect hints
of new physics beyond it. Kaon and pion physics are
obvious grounds where to perform such tests, for instance in the
well studied  $\pi_{l2}$ ($\pi\rightarrow l\nu_l$) and
$K_{l2}$ ($K\rightarrow l\nu_l$) decays, where $l= e$ or $\mu$.
Unfortunately, the relevance of these single decay channels in probing
the SM is severely hindered by our theoretical uncertainties
on non perturbative quantities like $f_{\pi}$ and $f_{K}$,
which still remain at the percent level.
On the other hand, in the ratios
$R_{\pi}\!=\!\Gamma(\pi\!\rightarrow\!e\nu)/\Gamma(\pi\!\rightarrow\! \mu\nu)$
and $R_{K}\!=\!\Gamma(K\!\rightarrow \!e\nu)/\Gamma(K\!\rightarrow\! \mu\nu)$
of the electronic and muonic modes,
the hadronic uncertainties cancel to a very large extent.
As a result, the SM predictions of $R_{\pi}$ and $R_{K}$ are known 
with excellent accuracy \cite{f} and this makes it possible to fully 
exploit the great experimental resolutions on $R_{\pi}$ \cite{pdg} 
and $R_{K}$ \cite{pdg,kl2_exp} to constrain new physics effects. 
Given our limited predictive power on  $f_{\pi}$ and $f_{K}$, deviations
from the $\mu-e$ universality represent the best hope we have at the moment
to detect new physics effects in $\pi_{l2}$ and $K_{l2}$.
The most recent NA48/2 result on $R_K$:
$$
R^{exp.}_{K}=(2.416\pm 0.043_{stat.} \pm 0.024_{syst.})
\cdot 10^{-5}\,\,\,\,\,\,\rm{NA48/2}
$$
which will further improve with current analysis, 
significantly improves on the previous PDG value, 
$R^{exp.}_{K}=(2.44\pm 0.11)\cdot 10^{-5}$.
This is to be compared with the SM prediction which reads:
$$
R^{SM}_{K}=(2.472\pm 0.001)\cdot 10^{-5}.
$$
Denoting by $\Delta r^{e-\mu}_{\!NP}$ the deviation from $\mu-e$ 
universality in $R_{K}$ due to new physics, i.e.:
\begin{equation}
\label{one}
R_{K}=\frac{\Gamma^{K\rightarrow e\nu_e}_{SM}}
{\Gamma^{K\rightarrow \mu\nu_\mu}_{SM}}
\left(1+\Delta r^{e-\mu}_{\!NP}\right),
\end{equation}
the NA48/2 result requires (at the $2\sigma$ level): 
\begin{equation}
-0.063\leq\Delta r^{e-\mu}_{\!NP}\leq 0.017 \,\,\,\,\,\,\rm{NA48/2}.
\end{equation}
In the following, we consider low-energy supersymmetric extensions 
of the SM (with R parity) as the source of new physics to be tested 
by $R_K$ \cite{kl2}.
The question we intend to address is whether 
SUSY can cause deviations from $\mu-e$ universality in $K_{l2}$ 
at a level which can be probed with the present attained experimental 
sensitivity, namely at the percent level.
We will show that i) it is indeed possible for regions of the minimal 
supersymmetric standard model (MSSM) to obtain $\Delta r^{e-\mu}_{\!NP}$ of
$\mathcal{O}(10^{-2})$ and ii) such large contributions to 
$K_{l2}$ do not arise from SUSY lepton flavor conserving (LFC) effects, but,
rather, from lepton flavor violating (LFV) ones.
Finally, being the NA48/2 $R_K$ central value below the SM prediction, 
one may wonder whether SUSY contributions could have the correct sign 
to account for such an effect.
We will show that there exist regions of the SUSY parameter space where 
the total $R_K$ arising from all such SM and SUSY terms is indeed lower 
than $R^{SM}_K$.

The SM contributions to $\pi_{l2}$ and $K_{l2}$ are helicity suppressed;
hence, these processes are very sensitive to non-SM effects.
In particular, charged Higgs bosons ($H^\pm$) appearing in any model with 
two Higgs doublets (including the SUSY case) can contribute at tree level to 
the above processes inducing the following effects \cite{hou}:
\begin{equation}
\label{tree}
\frac{\Gamma(M\!\rightarrow\! l\nu)}{\Gamma_{\!SM}(M\!\rightarrow\! l\nu)}=r_H=
\left[1-\left(\frac{m_d}{m_d+m_u}\right)^{2}\!\tan^{2}\!\beta
\frac{m^{2}_{M}}{m^{2}_{H}}\right]^2
\end{equation}
where $m_{u}$ is the mass of the up quark while $m_{s,d}$ stands for the
down-type quark mass of the $M$ meson ($M=K, \pi$).
From Eq.~(\ref{tree}) it is evident that such tree level contributions 
do not introduce any lepton flavour dependent correction.
The first SUSY contributions violating the $\mu-e$ universality in 
$M\rightarrow l\nu$ decays arise at the one-loop level with various diagrams 
involving exchanges of (charged and neutral) Higgs scalars, charginos, 
neutralinos and sleptons.
For our purpose, it is relevant to divide all such contributions into two classes:
i) LFC contributions where the charged meson M
decays without FCNC in the leptonic sector, i.e. $M\rightarrow l\nu_l$;
ii) LFV contributions $M\rightarrow l_i\nu_k$, with $i\neq k$
(in particular, the interesting case will be for $i= e,\mu$, and $k=\tau$).
A typical contribution of the first class is of order
\begin{equation}
\label{lfc1}
\Delta r^{e-\mu}_{SUSY}\sim \frac{\alpha_{2}}{4\pi}
\left(\!\frac{m^{2}_{\mu}-m^{2}_{e}}{m^{2}_{H}}\!\right)\tan^2\beta\,,
\end{equation}
where $H$ denotes a heavy Higgs circulating in the loop.
Then, even if we assume particularly favorable circumstances like
$\tan\beta=50$, we end up with  $\Delta r^{e-\mu}_{SUSY}\leq 10^{-6}$
much below the percent level of experimental sensitivity.
One could naively think that contributions of the second class
(LFV contributions) are further suppressed with respect to the LFC ones.
On the contrary, we show that charged Higgs mediated SUSY LFV
contributions, in particular in the kaon decays into an electron 
and a tau neutrino, can be strongly enhanced.
The quantity which now accounts for the deviation from the $\mu-e$
universality is $R^{LFV}_{\pi,K}=\sum_i\Gamma(\pi(K)\rightarrow e\nu_i)/
\sum_i\Gamma(\pi(K)\rightarrow \mu\nu_{i})$ (with $i= e,\mu,\tau$)
with the sum  extended over all (anti)neutrino flavors
(experimentally one determines only the charged lepton flavor in the decay products).
The dominant SUSY contributions to $R^{LFV}_{\pi,K}$ arise from the charged Higgs 
exchange. The effective LFV Yukawa couplings we consider are:
\begin{equation}
\label{coupl1}
l^{\mp}H^{\pm}{\nu_{\tau}}\rightarrow
\frac{g_2}{\sqrt
  2}\frac{{m_{\tau}}}{M_W}\Delta^{3l}_{R}\tan^{2}\!\beta
\,\,\,\,\,\,\,\,\,\,\,\,l=e,\mu.
\end{equation}
The $\Delta^{3l}_{R}$ terms are induced at one loop level by non holomorphic 
corrections through the exchange of gauginos and sleptons, provided LFV mixing 
among the sleptons \cite{bkl} 
(for phenomenological applications, see \cite{bkl,anna,ellis,hmio}).
Since the Yukawa operator is of dimension four, the quantities
$\Delta^{3l}_{R}$ depend only on ratios of SUSY masses, hence avoiding SUSY 
decoupling. $\Delta^{3l}_{R}$ is proportional to the off-diagonal
flavor changing entries of the slepton mass matrix
$\delta^{3j}_{RR}\!=\!({\tilde m}^2_{\ell})_{3_R j_R}/
\langle{\tilde m}^2_{\ell}\rangle$.
Following the thorough analysis in \cite{anna}, it turns out that
$\Delta^{3l}_{R}\leq 10^{-3}$.
Making use of the LFV Yukawa coupling in Eq.~(\ref{coupl1}),
it turns out that the dominant contribution to $\Delta r^{e-\mu}_{NP}$ 
reads \cite{kl2}:
\begin{equation}
\label{lfv}
R^{LFV}_{K}\simeq R^{SM}_{K}
\left[1+\left(\frac{m^{4}_{K}}{M^{4}_{H}}\right)
\!\left(\frac{m^{2}_{\tau}}{m^{2}_{e}}\right)|\Delta^{31}_{R}|^2\,
\tan^{\!6}\!\beta\right].
\end{equation}
Taking $\Delta^{31}_{R}\!\simeq\!5\cdot 10^{-4}$ accordingly to what said above,
$\tan\beta\!=\!40$ and $M_{H}\!=\!500 GeV$ we end up with
$R^{LFV}_{K}\!\simeq\!R^{SM}_{K}(1+0.013)$.
Turning to pion physics, one could wonder whether the analogous quantity
$\Delta r^{e-\mu}_{\!\pi\,SUSY}$ is able to constrain SUSY LFV. However, 
the correlation
$\Delta r^{e-\mu}_{\pi\,SUSY}\leq
(m^{4}_{\pi}/m^{4}_{k})\Delta r^{e-\mu}_{\!K\,SUSY}<10^{-4}$
clearly shows that the constraints on $\Delta r^{e-\mu}_{\!K\,susy}$ force
$\Delta r^{e-\mu}_{\pi\,susy}$ to be much below its actual experimental upper bound.
LFV effects to $\Delta r^{e-\mu}_{\!K\,SUSY}$ at the per cent level are allowed by 
the experimental bounds on LFV tau decays 
($Br(\tau\rightarrow l_j X)\leq 10^{-7}$, with $X=\gamma,\eta,\mu\mu$).
In fact, $\Delta r^{e-\mu}_{\!K\,SUSY}$ at the percent level corresponds to
$Br(\tau\rightarrow eX)\leq 10^{-10}$ \cite{hmio,1mio}.
The above SUSY dominant contribution to $\Delta r^{e-\mu}_{\!NP}$
increases the value of $R_{K}$ with respect to the SM expectation.
On the other hand, the recent NA48/2 result exhibits a
central value lower than $R_{K}^{SM}$.
One may wonder whether SUSY could account for such a lower $R_{K}$.
Obviously, the only way it can is through terms which, contributing to
the LFC $K\!\rightarrow\!l\nu_{l}$ channels, can interfere ( destructively)
with the SM contribution.
One can envisage the possibility of making use
of the large LFV contributions to give rise to LFC ones through double
LFV mass insertions in the scalar lepton propagators.

The corrections to the LFC $H^{\pm}l\nu_{l}$ vertices induced by LFV effects are:
\begin{equation}
\label{coupl2}
l^{\mp}H^{\pm}{\nu_{l}}\rightarrow 
\frac{g_2}{\sqrt 2}\frac{{m_{l}}}{M_W}\tan\!\beta
\left(1+\frac{m_{\tau}}{m_{l}}\Delta^{ll}_{RL}
\tan\!\beta\right)\,\,\,\,\,\,l=e,\mu
\end{equation}
where the second term is generated by a double LFV source
that, as a final effect, preserves the flavour. Indeed $\Delta^{ll}_{RL}$
is proportional to $\delta^{l3}_{RR}\delta^{3l}_{LL}$.
In the large slepton mixing case, $\Delta^{ll}_{RL}$ terms are of the
same order of $\Delta^{3l}_{R}$.
These new effects modify the previous $R^{LFV}_{K}$ expression in the 
following way \cite{kl2}:
\begin{equation}
\label{lfclfv}
R^{LFV}_{K}\simeq R^{SM}_{K}\,
\bigg[\,\bigg|1\!-\!\frac{m^{2}_{K}}{M^{2}_{H}}
\frac{m_{\tau}}{m_{e}}\Delta^{11}_{RL}\,\tan^{\!3}\!\beta\,\bigg|^{2}\!+\!
\bigg(\frac{m^{4}_{K}}{M^{4}_{H}}\bigg)
\!\bigg(\frac{m^{2}_{\tau}}{m^{2}_{e}}\bigg)
|\Delta^{31}_{R}|^2\,\tan^{\!6}\!\beta
\bigg].
\end{equation}
Setting the parameters as in the example of the above section and if
$\Delta^{11}_{RL}\!=\!10^{-4}$ we get $R^{LFV}_{K}\!\simeq\! R^{SM}_{K}(1-0.032)$.

In the most favorable scenarios, the deviations from the SM could reach 
$\sim 1\%$ in the $R_K^{\mu/e}$ case \cite{kl2} (not far from the present 
experimental resolution \cite{kl2_exp}) 
and $\sim {\rm few} \times 10^{-4}$ in the $R_\pi^{\mu/e}$
case. In the pion case the effect is quite below the 
present experimental resolution~\cite{pl2_exp}, 
but could well be within the reach of the new
generation of high-precision $\pi_{\ell 2}$ 
experiments planned at TRIUMPH and at PSI.
  
In principle, larger violations of LF universality are expected 
in $B \to \ell \nu $ decays, with $\cO(10\%)$ deviations 
from the SM in $R_B^{\mu/\tau}$ and even order-of-magnitude 
enhancements in $R_B^{e/\tau}$~\cite{tgbhints}.
However, the difficulty of precision measurements 
of the highly suppressed $B \to e/\mu~ \nu$ modes
makes these non-standard effects undetectable (at least at present). 

Similarly to the FCNC decays, also for the LF universality tests 
the low-energy systems ($K_{\ell 2}$ and $\pi_{\ell 2}$)
offer a unique opportunity in shedding light on physics beyond 
the Standard Model: the smallness of NP effects is more than compensated 
(in terms of NP sensitivity) by the excellent experimental resolution and 
the good theoretical control.

Finally, we remark that a key ingredient of all the effects discussed in 
the present section are large $\tan\beta$ values so, it is legitimate to ask 
how natural is this framework.
The regime of large $\tan\beta$ [$\tan\beta = \mathcal(m_t/m_b)$] has an 
intrinsic theoretical interest since it allows the unification of 
top and bottom Yukawa couplings, as predicted 
in well-motivated grand-unified models.
Moreover, as recently discussed in \cite{tgbhints},
this scenario is particularly appealing also from a phenomenological
point of view.
In fact, in this framework, one could naturally accommodate 
the present central values of both $BR(B\rightarrow \tau\nu)$ and $(g-2)_\mu$, 
explain why the lightest Higgs boson has not been observed yet,
and why no signal of new physics has been observed in $BR(B\to X_s \gamma)$ 
and $\Delta M_{B_s}$ without requiring any fine tuning.
So, one of the virtues of the large $\tan\beta$ regime of the 
MSSM is its naturalness in flavor physics and in precise 
electroweak tests.

\end{document}